\documentclass[aps,prl,twocolumn,groupedaddress,floatfix,showpacs]{revtex4}
\usepackage{amssymb,amsmath}
\usepackage{graphicx}
\usepackage{subfigure}
\usepackage[english]{babel}
\usepackage{float}
\usepackage{color}

\begin{document}

\title{Observation of bound states in Lieb photonic lattices}

\author{Rodrigo A. Vicencio$^{1}$, Camilo Cantillano$^{1}$, Luis Morales-Inostroza$^{1}$, Basti\'an Real$^{1}$, Steffen Weimann$^{2}$, Alexander Szameit$^{2}$, and Mario I. Molina$^{1}$, }

\address{$^1$Departamento de F\'{\i}sica, MSI-Nucleus on Advanced Optics, and Center for Optics and Photonics (CEFOP), Facultad de Ciencias, Universidad de Chile, Santiago, Chile}

\affiliation{$^{2}$Institute of Applied Physics, Friedrich-Schiller-Universit\"at Jena, Max-Wien-Platz 1, 07743 Jena, Germany}

\pacs{63.20.Pw, 42.82.Et, 78.67.Pt}

\begin{abstract}
We present the first experimental demonstration of a new type of bound states in the continuum, namely, compacton-like linear states in flat bands lattices. To this end, photonic Lieb lattices are employed, which exhibit three tight-binding bands, with one being perfectly flat. Our results could be of great importance for fundamental physics as well as for various applications concerning imaging and data transmission.
\end{abstract}

\maketitle

Localization of excitations in periodical lattices usually have to rely on some form of defects either point-like or extended, linear or nonlinear
or simply disorder, in order to produce localized modes. However, there is another way in which the lattice remains perfectly periodic but is able to
support localized states.  In that case, one can find extremely localized
entities that do not diffract at all, and remain localized by virtue of a perfect geometric phase cancellation condition. In an optical context,
this implies the possibility of using a judicious combination of these modes to transmit information along waveguide arrays without any distorsion. Here,
we demonstrate theoretically and experimentally the existence of such compacton-like linear states residing in the linear band of a Lieb
lattice. Most importantly, they constitute a new type of bound state in the continuum.

A Lieb lattice [see Fig.1(a)] is a square depleted lattice that is essentially a two-dimensional counterpart of the ``perovskite'' structure, which
is ubiquitous in nature. The $CuO_{2}$ planes of cuprate superconductors are perhaps the most famous example~\cite{cero_a, cero_b}, but in fact a lot
of layered oxides coordinate in this fashion. Initial interest on this lattice started when ferromagnetism was found on the flat band at half
filling~\cite{uno}. Later, it was proven that ferromagnetism in this lattice was robust against spin wave excitations~\cite{dos}. This lattice also
display unusual topological properties. For instance, in the presence of a uniform magnetic field, the flat band remains flat because of topological
reasons~\cite{tres}. The flat band touches two linearly dispersing intersecting bands at a single Dirac point. In the presence of Kerr nonlinearity
the system may exhibit novel conical diffraction at the Dirac cone~\cite{anton,lieb2}. Very recently, it was studied the effect of considering correlated disorder on a Lieb lattice~\cite{flach1}, where a square root singularity in the density of states was predicted.

The presence of a flat band in the spectrum of a Lieb lattice implies the existence of entirely degenerate states, whose superposition displays no dynamical evolution. This allows the formation of four-site ring structures that are completely localized and constitute a new type of bound state in the continuum~\cite{bic1,bic2,bic3}. The Lieb lattice can be realized by, e.g., manipulating cold atoms in optical lattices~\cite{cuatro_a,cuatro_b,cuatro_c} and by direct laser-writing of optical
waveguides~\cite{lieb1}. In this work we will take advantage of the flat band of the Lieb lattice and build experimentally compacton-like localized
entities that do not diffract upon propagation, allowing in this manner the distorsionless transmission of information along an optical channel~\cite{kagome}.

\begin{figure}[t]
\begin{center}
\includegraphics[width=0.49\textwidth]{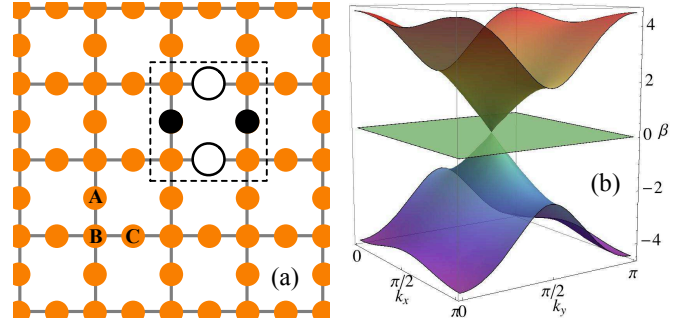}
\caption{(a) Transversal plane of a Lieb lattice with sites $A$, $B$ and $C$ defining the unitary cell. The dashed-line region encloses a ring mode
profile. (b) Linear spectrum of an anisotropic Lieb lattice for $V_y=2V_x$.}\label{fig1}
\end{center}
\end{figure}

The evolution of light along the $z$ direction on a Lieb photonic lattice, sketched in Fig.~\ref{fig1}(a), composed by weakly-coupled identical
optical waveguides, is well described by a discrete linear Schr\"{o}dinger equation~\cite{rep1,rep2}:
\begin{equation}
-i\frac{d u_{\vec{n}}}{d z} = \beta_{\vec{n}} u_{\vec{n}}+\sum_{\vec{m}} V_{\vec{n},\vec{m}}u_{\vec{m}}\ .
\label{model}
\end{equation}
Here, $z$ is the coordinate along the propagation direction, and $u_{\vec{n}}$ corresponds to the light amplitude at the $\vec{n}$-th
%
%
waveguide of a Lieb lattice. In our model, all waveguides are assumed to be identical, therefore they possess the same propagation constants
$\beta_{\vec{n}}=\beta_0$ [without loss of generality, we set $\beta_0=0$ in  (\ref{model})]. The coupling (hopping) term between
nearest-neighbor lattice sites $\vec{n}$ and $\vec{m}$ is given by $V_{\vec{n},\vec{m}}$. The linear spectrum of a Lieb lattice is obtained by
solving model (\ref{model}) with a stationary ansatz: $u_{\vec{n}}(z)=u_{\vec{n}}\exp(i \beta z)$. We assume nearest-neighbor interactions only
between sites $A$, $B$ and $C$ (unitary cell), as shown in Fig.~\ref{fig1}(a). To be more general, we consider an anisotropic lattice where the
horizontal ($V_x$) and vertical ($V_y$) coupling coefficients can be different, but preserve the same lattice properties (diagonal interactions are
considered negligible in our approach). For these three different lattice sites, we construct the coupling interactions considering a 2D Bloch
wavevector $\vec{k}=\{k_x,k_y\}$. We find three linear bands on this system~\cite{lieb1}:
$$\beta(\vec{k})=0\ ,\ \ \pm 2 \sqrt{V_x^2\cos^2(k_x) +V_y^2\cos^2(k_y)}\ .$$
In Fig.~\ref{fig1}(b), the linear spectrum is shown inside the first Brillouin zone. Two bands are dispersive (nonzero curvature), possessing a
particle-hole symmetry~\cite{phs}, where for each $\vec{k}$ there are two eigenfrequencies $\pm\beta(\vec{k})$. We observe that for $V_y>V_x$, the
dispersive bands show a different curvature depending on the direction of the $\vec{k}$-vector, where $y$-oriented waves propagate, in general,
faster. These two bands are connected by a Dirac point at $\beta=0$. Exactly at this value, a completely flat (non-dispersive) band is located. 

In general, modes in the continuum of any periodic structure are completely extended. However, a flat band system allows the formation of very localized, compacton-like, bound states~\cite{berg,flach1}. 
In a Lieb lattice, any closed ring (formed by eight sites) may support a ring mode, where $B$-amplitudes are zero and the other two amplitudes satisfy the
relation: $V_x C =-V_y A$, for $k_y=k_x$, as sketched in Fig.~\ref{fig1}(a) [see dashed line]. A measure of localization is provided by the
participation ratio $R=(\sum_{\bf n} |u_{\bf n}|^2)^2/\sum_{\bf n} |u_{\bf n}|^4$, which takes the value $1$ for a state localized at a single site and $N$
for a completely delocalized profile (being $N$ the number of lattice sites). In our case, the ring modes have a participation ratio  $\leqslant4$; i.e., they constitute a very compact
\textit{bound state embedded into the continuum (BIC)}~\cite{bic1,bic2,bic3}, appearing only by virtue of symmetry in a completely periodic lattice. This BIC can be located in any position across the lattice and
will propagate without diffraction along the longitudinal direction. Additionally, any linear combination of them will be completely coherent and
will propagate without any distortion, allowing for a high-fidelity transmission of information along the longitudinal direction~\cite{kagome}.

Tight-binding models, for example model (\ref{model}), are known to describe qualitatively well some particular systems having short-range (weak
coupling) interactions~\cite{rep1,rep2}. However, the description of a real lattice require other methods to study the dynamics of light propagating. Our goal in this work is to trace the optimal conditions for observing the discrete phenomenology, avoiding extra interactions
that actually destroy the flatness of the band~\cite{desa} and, therefore, do not allow the observation of linear compact states. In order to be
closer to the conditions likely to be found in an experiment~\cite{lieb1}, we study dynamically a Lieb lattice using a continuous approach. We consider a linear
paraxial wave-equation given by
\begin{equation}
-i \frac{\partial}{\partial z} \psi(x,y,z) = \frac{\nabla_{\bot}^2 \psi(x,y,z)}{2k_0n_0} + k_0\Delta n(x,y)\psi(x,y,z)\ ,
\label{model2}
\end{equation}
where $\psi$ is the envelope of the electric field, $k_0=2\pi/\lambda$ is the wavenumber in free space, $\lambda$ is the wavelength, $n_0$ is the
refractive index of the bulk material, and $\Delta n(x,y)$ corresponds to the refractive index structure which defines the Lieb photonic lattice.
$\nabla_{\bot}^2 =\partial_x^2+\partial_y^2$ corresponds to the transverse Laplacian operator. Once the geometry is fixed [in our case, waveguides
are elliptical~\cite{lieb1}, implying a strong effective anisotropy: $V_y\sim 2V_x$ in model $(\ref{model})$], we have essentially two free parameters left, the wavelength
and the maximum index contrast $\delta n\equiv \mbox{Max}|\Delta n(x,y)-n_{0}|$.
%
\begin{figure}[t]
\begin{center}
\includegraphics[width=0.47\textwidth]{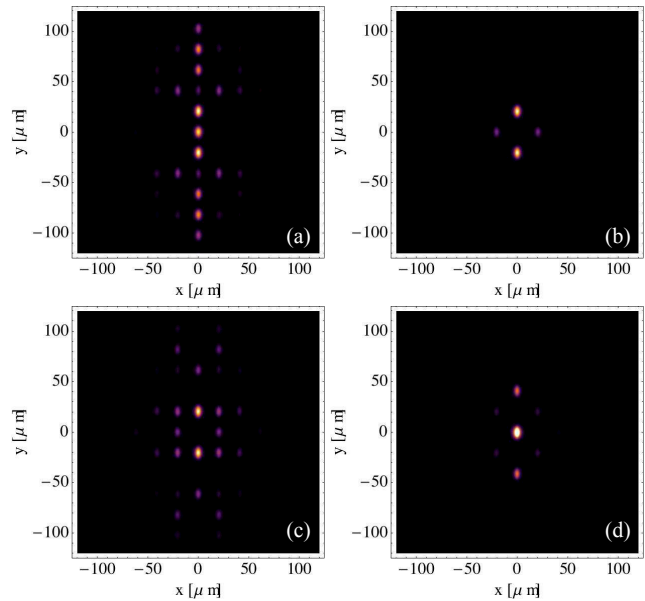}
\caption{Output profiles for (a) bulk B-site, (b) ring mode, (c) four sites in phase, and (d) two added ring modes excitations. $\lambda=532$ nm,
$d=20\ \mu$m, $\delta n=0.7\times10^{-3}$, and $L=10$ cm.}\label{fig2}
\end{center}
\end{figure}
%
We implement a Beam Propagation Method to solve Eq.(\ref{model2}) numerically and study the evolution of the light across and along a Lieb lattice,
with a lattice period of $d=20\ \mu$m and a propagation distance of $L=10$ cm. In our first experiment we used light 
with $\lambda=633$ nm, and observed a well diffracted pattern when exciting a bulk $B$-site. However, green light of
$\lambda=532$ nm showed less diffraction~\cite{lieb2} and, therefore, was closer to a discrete phenomenology. By running our
simulations at this wavelength, we obtained the results presented in Fig.~\ref{fig2}. First of all, in Fig.~\ref{fig2}(a) we observe a dramatic
reduction of the diffraction area just by changing the wavelength (when comparing with recent results~\cite{lieb1}), which is a manifestation of a
more discrete (weak-interaction) scenario. In addition, due to the waveguide ellipticity, we observe a very anisotropic and vertically oriented
dispersion pattern, which can be interpreted as $V_y>V_x$ in the discrete model. In Fig.~\ref{fig2}(b), we observe how a ring mode could effectively be excited by choosing a right
index contrast $\delta n$ (we ran several simulations to find that $\delta n\gtrsim0.67\times10^{-3}$). In Fig.~\ref{fig2}(c) we observe how a ring profile with a zero phase structure is destroyed and diffract over the lattice. This shows the importance of given the right phase structure ($\pi$) to effectively excite the flat band bound states. Finally, as a proof of the numerical excitation of the ring mode,
we generate a coherently combined state, by summing two neighboring ring modes, and propagating it along the system. In Fig.~\ref{fig2}(d) we
observe the output pattern of this combined state, showing a perfect propagation of ring mode combinations. We generated different, simple and complex, linear
combinations and all of them propagated without noticeable distortion. 

The excitation of these localized states gives also rise to the onset of different excited eigenstates coming from the complete band structure of the lattice; in our numerical simulations, this manifests as a very weak (negligible) background radiation. We performed a longitudinal Fourier Transform analysis~\cite{kagome} and study the excited frequency spectrum along the dynamics. We observed the presence of a large and thin peak in the region of the flat band, with a lower excitation of higher bands. By increasing the index contrast $\delta n$, we notice that this flat band peak increases as well as the gap in between the nearest excited bands. Additionally, we performed simulations for propagation distances up to $L=50$ cm to study the robustness of the ring mode excitation. By inspecting the excited spectrum we observed that the peak related to flat band remained strongly excited in comparison to the rest of the excited spectrum. Although a realistic excitation of a ring mode is not perfect as it is in a discrete model, we numerically found that the generated background radiation is very weak and can be considered negligible in comparison to the excited bound state.

\begin{figure}[t]
\begin{center}
\includegraphics[width=0.47\textwidth]{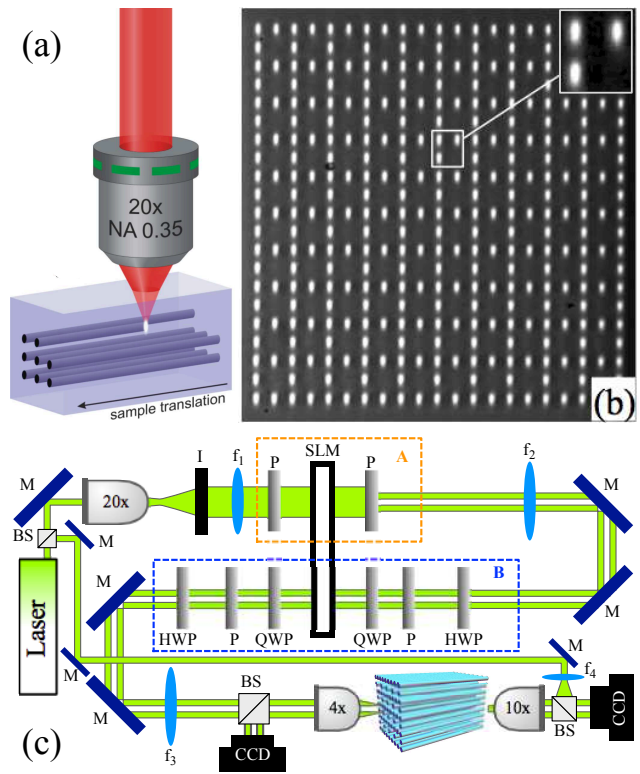}
\caption{(a) Femtosecond-laser writing technique. (b) Microscope image at the output facet of a Lieb lattice for white light propagation. (c)
Experimental setup to study the propagation of light patterns on a Lieb lattice.}\label{fig3}
\end{center}
\end{figure}

To perform the experiments, we fabricate a Lieb photonic lattice using the femtosecond-laser technique on a $L=10$ cm long fused silica glass
wafer~\cite{fslt}, as sketched in Fig.~\ref{fig3}(a). In order to test the quality of this lattice, we launch white light at the input facet and take
a microscope image at the output of the crystal [see Fig.~\ref{fig3}(b)]. In this figure, we observe the propagating modes of each waveguide, showing a noticeable ellipticity that strongly affects the coupling interactions between nearest-neighbors. We observe more evanescent light in between vertical sites than in between horizontal ones [see zoom in Fig.~\ref{fig3}(b)], due to the effective anisotropic coupling ($V_y>V_x$).
%
%

The experimental observation of ring modes involves several challenging stages for the preparation of a proper initial condition.
We essentially require to create an amplitude and phase pattern to be used as an input excitation, at the input facet of the array. In
Fig.~\ref{fig3}(c) we describe our experimental setup. We use a Holoeye LC2012 transmission spatial light modulator (SLM) to modulate,
simultaneously, the amplitude and phase of a broad beam. We split the SLM display into two parts. In the first path (region $A$), we
modulate the amplitude of the beam by tuning the angle of two polarizers (P). We generate a pattern of several light disks of given radius (to be
adjusted to match waveguides at the input facet) and given geometry (considering the Lieb structure). Once this modulation is achieved, we pass this
light pattern through the second part of the SLM to modulate now the amplitude profile in phase ($0$ or $\pi$). This modulation is performed in
region $B$, by using an array of tuned half-wave (HWP) and quarter-wave (QWP) plates, and polarizers (P) (all components were carefully tuned to
optimize the required phase modulation, as well as the intensity of the profile, and the respective input polarization~\cite{slm}). After this stage, we obtain
an amplitude and phase modulated light pattern, with a given polarization (in our experiment, light is polarized in the horizontal direction, in order to
observe a larger diffraction area~\cite{poladiso}). Finally, we inject this modulated pattern at the input facet of our Lieb photonic lattice by using
a $4\times$ microscope objective (MO), and observe the input profile with a CCD camera after being reflected on a beam splitter (BS). We obtain the
output profile by using a $10\times$ MO and a CCD camera. To study the phase profile, we interfere the output and the input pattern with a wide
tilted plane wave, obtained by splitting the output of the laser with a BS.
%
\begin{figure}[t]
\begin{center}
\includegraphics[width=0.48\textwidth]{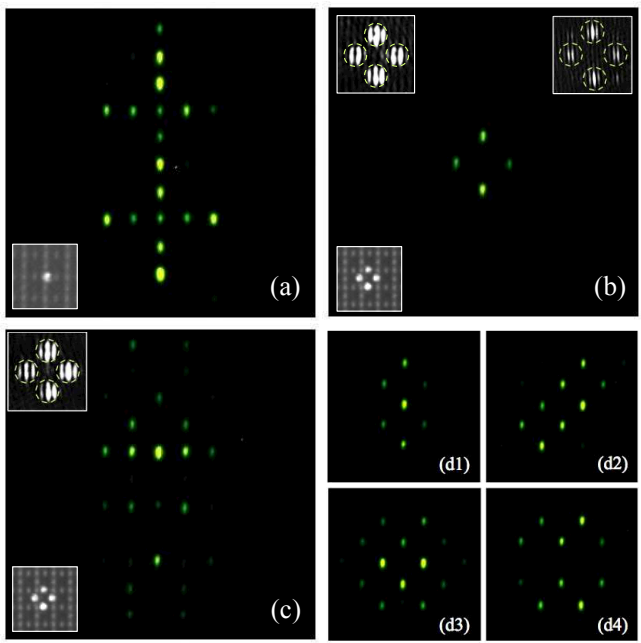}
\caption{Output experimental patterns for (a) bulk B-site, (b) ring mode, (c) four sites in phase, and (d) composed states excitations. In (a)--(c)
left-down insets show the input intensity profile. Upper insets in (b) and (c) show an interferogram of the ring mode with a tilted
plane wave for input (left) and output (right) profiles.}\label{fig4}
\end{center}
\end{figure}
%

First, we study the bulk diffraction by using a B-site excitation, which is the input condition that better excite the dispersive part of the linear spectrum~\cite{lieb1}. We inject light at the middle of the lattice and observe the diffraction
pattern shown in Fig.~\ref{fig4}(a). We found an excellent agreement with respect to our numerical results [see Fig.\ref{fig2}(a)], which is very
important for the calibration of the effective index contrast to be used in our simulations. Then, we prepare a symmetric ring mode input profile, having four
light disks with equal amplitude and a staggered phase structure, and observe its  perfect propagation along the lattice [see Fig.~\ref{fig4}(b)]. As the input profile is close but not the exact ring mode of the system, the light relaxes to this non symmetric configuration due to the anisotropic coupling coefficients. By increasing the image contrast (not shown here), we observed that a very weak radiation is also generated across the lattice, what is a manifestation of the presence of the dispersive bands in the system. By interfering our input and output ring mode profiles with a tilted plane wave, we observe that the phase structure is exactly
the one expected [see Fig.~\ref{fig4}(b)-inset]: there is a $\pi$-phase shift between neighboring sites as predicted by the tight-binding model. This shows very nicely that the
initial discrete model prediction is valid in a realistic environment, and that the fundamental properties of the Lieb lattice are really observable in this experiment. We test the relevance of the phase structure on the input condition, by using an in-phase four site ring excitation as shown in Fig.~\ref{fig4}(c)-insets. The output pattern shows a destruction of the initial localized profile due to the excitation of different linear bands (the output phase profile is not shown). This experiment agrees well with our simulations shown in Fig.~\ref{fig2}(c), with a main difference on the symmetry of the observed profile, what is essentially due to intrinsic fabrication anisotropies or from not perfectly alligned input conditions).

Finally, in order to probe the excitation of ring modes as stationary states and its potential use for transmitting optical information, we combine them in different
configurations. First of all, we confirm our numerics from Fig.~\ref{fig2}(d), by preparing an initial condition composed by two vertically added
ring modes and observing its almost perfect propagation in Fig.~\ref{fig4}(d1). This shows the robustness of the application of the discrete analysis in our experiments.
This pattern has no discernible background and possesses the same predicted structure for the combination of two anisotropic ring modes.
Additionally, the vertical combination was performed thinking on the highest possible vertical diffraction to be excited, nevertheless we observed a
perfect localized propagation. Finally, we tested other linear combinations. For example, in Fig.~\ref{fig4}(d2) we observe the propagation of two
non-interacting ring modes located along a diagonal. We also studied the propagation of more complex patterns by linearly combining four ring modes. We constructed and propagated a completely additive combination [Fig.~\ref{fig4}(d3)], and a horizontally-added and vertically-subtracted one [Fig.~\ref{fig4}(d4)]. By carefully selecting the
phase and location of these modes, different linear patterns can be formed to create a code based on these highly localized fundamental bound states.

In conclusion, we have experimentally observed, for the first time ever, the excitation of a bound state into the continuum of a Lieb photonic
lattice. We have matched the right realistic conditions to observe this very fundamental weak-coupling mode and demonstrate the reality of flat band systems. We have given a proof for the need of having a correct phase structure to correctly excite these very localized
entities. Additionally, we have combined these modes to create composite coherent and localized states, that propagate without diffraction across the lattice. This
shows the possibility to create different patterns and propagate them as a secure image transmission mechanism. Our results show a new way to
propagate localized patterns by using the very fundamental properties of flat band systems that can be found and studied in a broad class of physical systems,
ranging from solid-state and magnetism to photonics.

The authors wish to thank A. Desyatnikov and C. Mej\'ia-Cort\'es for valuable discussions. This work was supported in part by Fondef IDeA CA13I10244, Fondecyt grant 1120123, Programa ICM P10-030-F, the Programa de
Financiamiento Basal de CONICYT (FB0824/2008), the Deutsche Forschungsgemeinschaft (grant NO 462/6-1), and the German Ministry of Education and Research (Center for Innovation Competence program, grant 03Z1HN31).

\end{document}